# INTERGRANULAR FLUX PINNING IN UNDERDOPED AND OVERDOPED $R_{1-x}Ca_xBa_2Cu_3O_z$ (R=Y,Gd; x=0, 0.2) SAMPLES

E. K. Nazarova[1*], A. J. Zaleski[2], K. A. Nenkov[3,4] and A. L. Zahariev[1]

[1] *Georgi Nadjakov Institute of Solid State Physics, Bulgarian Academy of Sciences, 1784 Sofia, Bulgaria*
[2] *Institute of Low Temperature and Structure Research, PAS, 50-950 Wroclaw, Poland*
[3] *IFW, Leibniz Institute for Solid State and Materials Research, P.O. BOX 270016, D-01171 Dresden, Germany*
[4] *International Laboratory of High Magnetic Fields and Low Temperatures, 53-421 Wroclaw, Poland*

## Abstract

The temperature dependences of AC magnetic susceptibility at different magnetic field amplitudes and frequencies are investigated for underdoped and overdoped $R_{(1-x)}Ca_xBa_2Cu_3O_{7-\delta}$ (R=Y; Gd and x=0; 0.2) polycrystalline samples. The activation energy, $E_a$, for thermally assisted flux flow (TAFF) in intergranular region is determined. It was established that a correlation exists between the intergranular critical current and flux pinning activation energy. In underdoped samples the intergranular current shows S-I-S behaviour and the activation energy is small, while in overdoped samples the intergranular current is changed to S-N-S type and the activation energy increases. 2D pancake vortices are characteristic of underdoped samples, while 3D vortex system exists in overdoped samples. In fact we demonstrate that Ca substitution not only increases carrier concentration, but improves intergranular activation energy for TAFF.

[*]Corresponding author: Tel.: +359 2 9795000-211; Fax: +359 2 975 36 32
E-mail address: nazarova@issp.bas.bg

**Introduction**

In spite of their widely varying structure (different number of $CuO_2$ planes) high temperature superconductors (HTS) have a universal phase behavior. The scaled transition temperature, $T_c/T_{cmax}$ is a parabolic function of the plane hole concentration (p) with onset, maximum and termination for p=0.05, 0.16 and 0.27, respectively [1]. Underdoped 123-type samples can be obtained by a variation of oxygen content in the range of $6.3 \leq z \leq 6.92$. In the underdoped region all HTS have the same $n_s/m^*$ values ($n_s$ – superconducting carrier concentration and $m^*$ - effective mass) for the same $T_c$ [2], in accordance with Uemura relation. The optimal doping at p=0.16 is connected with maximal $T_c$, while at critical doping (p=0.19) the superconductivity is the most robust due to maximization of the condensation energy and superfluid density (at T=0) [3]. Oxygen non-deficient $YBa_2Cu_3O_7$ is slightly overdoped [4], while for the obtaining of heavily overdoped 123 superconductors an additional holes supply is needed. Partial replacement of rare earth ion $R^{3+}$ by $Ca^{2+}$ ion, with similar ionic radius but lower valence value, provides additional holes and makes the overdoped region of the phase diagram accessible for study. It is surprising, however, that the increasing of the charge concentration in the plane leads to decreasing of $T_c$ on the overdoped side. The reduction of $T_c$ is explained by the increasing of the 3D electron dynamics when superconductivity can be strongly suppressed by electron-electron scattering and scattering on structural disorder (M-O interlayer between the $CuO_2$ sheets) [5]. In the presence of nonmagnetic substituents (like Zn on the Cu site and Ca on the R site) the impurities are also strongly scattering centers. Zn, for example, induces local magnetic moments upon its neighboring copper atoms [6] and the spin vacancy creates a perturbation of the local antiferromagnetic correlations [7] The induced magnetic moment decreases with increasing the doping, but it has been detected even in Ca substituted overdoped YBCO. However from the viewpoint of intragrain pinning these magnetic perturbations are effective pinning centers. Y. Zhao et al. [8] show that Ni doped samples exhibit higher pinning than Y123 samples with Y211 particles at fields higher than 2T. A better understanding of the interconnection between overdoping and $T_c$ suppression is of importance. By a control of the fine scale microstructure (by substitution, for example) it is possible to engineer superconducting materials with different plane carrier concentrations and superconducting properties.

Chemical doping with Ca is important also with respect to the intergranular current. It is known that grain boundaries (GBs) in YBCO superconductors are depleted of carriers [9, 10], what leads to a reduced intergrain critical current density. Schmehl et al. [11] reported that by substituting $Ca^{2+}$ at $Y^{3+}$ site, the weak link effect of high angle GBs in YBCO multilayer films was significantly reduced and the critical current density was enhanced more than seven times. An improvement of GBs transport in bulk melt-processed [12] and sintered [13] YBCO samples by a Ca substitution has been also established. The intergranular current is important not only by itself, but as an element governing the flux pinning and activation energy for flux creep [14, 15].

Hole concentration is an essential parameter that controls many properties in HTS. How the vortex dynamics is influenced by the doping effect? Samples belonging to underdoped region have lower charge concentration in plane *p*, so almost no "chain condensate" exists. In the overdoped side the charge concentration in planes increases, 3D carrier dynamics is enhanced and the anisotropy decreases. Generally, doping strongly affects coupling between the $CuO_2$ layers influencing the vortex dynamics in grains. In this study we investigate how the doping influences the intergranular flux dynamics by comparing the properties of the samples from the opposite sides of the T(p) phase diagram.

**Experimental**

The investigated polycrystalline samples are $R_{1-x}Ca_xBa_2Cu_3O_{7-\delta}$ where R=Y, Gd and x=0; 0.2. Several tablets have been prepared from the composition $Y_{0.8}Ca_{0.2}Ba_2Cu_3O_{7-\delta}$. One of the tablets is synthesized in nitrogen atmosphere in order to obtain fully deoxygenated sample (Y0.8-N) with carriers supplied only by Ca. In order to prepare highly overdoped samples, one of the $Y_{0.8}Ca_{0.2}Ba_2Cu_3O_{7-\delta}$ tablets was oxygenated for 48 hours at 320 °C (Y08-48) and the other at 450 °C for about 100 hours (Y0.8-100). Both samples Gd123 and Y123 (with x=0) are obtained by standard solid-state reaction method, synthesized in oxygen atmosphere at temperature 950 °C for 23 hours and subsequently annealed at 450 °C for another 23 hours and furnace cooled to room temperature.

Samples were analyzed by X-ray diffraction ($Cu_{k\alpha}$ radiation) and AC magnetic susceptibility measurements. The latter are performed with MagLab-Oxford 7000 susceptometer for different AC field amplitudes, frequencies and temperatures.

**Results and Discussion**

A. Samples

The investigated samples have different carrier concentrations and represent both underdoped and overdoped sides of the T(p) phase diagram. A complete list of studied specimens is given in Table 1.

According to XRD analysis fully deoxygenated Ca-doped compound $Y_{0.8}Ca_{0.2}Ba_2Cu_3O_z$ –(denoted as Y0.8-N) has tetragonal crystal structure. It has been indicated by the following peaks: (103), (110), (200), (006), (213) and

(116) from the diffractogram presented on Fig.1. Lattice parameters of tetragonal structure are a=3.849 A and c=11.818 A (see Table 1). Using the correlation between the **c** lattice parameter and the oxygen content in Ca substituted samples [16, 17], we estimate the oxygen content in Y08-N sample to be 6.0÷6.2. In fact the sample consists of two types tetragonal unit cells: some of them Ca substituted, others non substituted but both with out chain oxygen. Thus, the number of carriers per $CuO_2$ plane is determined only from Ca concentration (p=x/2=0.1), i.e. the sample is highly underdoped. Its critical temperature, determined as the onset of χ'(T) dependence, is between 40 and 50 K (Fig.2a) and also confirms the underdoping.

The large deviation from $T_{c\;max}$ is observed for Gd00 sample (about 2.3 K), which also has a rather broad transition width. This is usually connected with the formation of R-enriched ($R_{1+y}Ba_{2-y}Cu_3O_z$) clusters during the synthesis in oxygen atmosphere [18]. $T_c$ decreases due to the reduction of carrier concentration when R ions with $3^+$ valence substitute for Ba with $2^+$ valence. According to XRD analysis the sample is in single phase with orthorhombic crystal structure. Even when the $T_{c,onset}$ is high, a broad transition is usually observed in the AC susceptibility vs. temperature dependence. It is reasonable to assume that $T_c$ suppression in Gd00 sample is rather a manifestation of underdoping than of overdoping. Using the experimental data from [19], we estimated $T_{cmax}$ to be 94.0 K for $GdBa_2Cu_3O_{7-\delta}$ and the calculated **p** value is 0.133.

The Y123 sample is orthorhombic with $T_c$=92.0 K (Table 1). It is slightly overdoped with **p**=0.189 determined from $T_c/T_{cmac}=1-82.6\;(p-0.16)^2$ empirical dependence, where $T_{cmax}$=93.5 K.

$T_c$ decreases in all $RBa_2Cu_3O_x$ compounds when Ca substitution takes place into R position (Table 1). Both substituted samples (Y08-48, Y08-100) are long time oxygenated with small **c** lattice parameter of order of 11. 666 A (see Table 1). Lattice parameter **a** is almost unchanged, while **b** is higher for the longer oxygenated sample. Their intragranular critical temperatures are 82.7 K and 81.5

K, respectively. These values are well below the $T_{cmax}$, which is found to be about 85 K for this compound [19]. This is an indication for overdoping of both samples. $T_{c,intra}$ and $T_{c,inter}$ are determined from AC susceptibility measurements ($H_{ac}$=7.95 a/m (0.1 Oe) and f=145 Hz) and are presented in Table 1.

B. Intergranular flux pinning

AC magnetic susceptibility is a powerful tool to study the flux lines dynamics in high temperature superconductors. AC technique allows inducing changes in the vortex dynamics by variations of AC field amplitude and frequency, temperature and DC field. The peak in the imaginary part of the AC susceptibility, $\chi''_1(T)$, occurs when the measuring frequency is of the order of the inverse relaxation time of the vortex system [20]. The peak position depends logarithmically on AC field frequency. This dependence is used for the determination of activation energy, $E_a$, needed for starting the flux creep, when thermally activated fluctuations overcome the pinning [21]. The expression for $E_a$ is given by:

$$E_a = kT_p \ln(f/f_0),$$

where k is the Boltzmann constant. $T_p$ is the peak temperature (corresponding to the $\chi''_1$ peak position) and $f_0$ is a characteristic frequency in the range $10^9$ – $10^{12}$ Hz.

It has been established that in all investigated samples intergranular $\chi''_1(T)$ dependences are shifted to higher temperatures when the frequency is increased at constant AC field amplitude. The peaks of $\chi''_1(T)$ dependences are shifted to lower temperatures when the magnetic field amplitude increases at constant low frequency. These experimental results indicate that intergranular flux dynamics is influenced by the thermally activated flux creep. This is in agreement with the previous reports [14]. Therefore, based on the above equation, we have calculated the flux creep activation energies for all examined samples.

The sample Y08-N has a very low $T_c$ and the maxima in $\chi''_1(T)$ dependences are not well expressed (Fig.2b) down to 2K even for amplitude $H_{ac}$=7.95 a/m (0.1 Oe). For $E_a$ determination we used the curves displacement with frequency on the level possibly very close to the maximum. We chose this procedure after a careful investigation of the other samples where the maximum of $\chi''_1(T)$ is well manifested. Once the $1/T_p$ vs. ln(**f**) dependence has been built taking the values of $T_p$ at the maximum, then the same dependence is made but for $T_p$, we took the value at which the line y=n [max $\chi''_1(T)$], n=0.7 ÷ 0.9 intercepts the $\chi''_1(T)$ plots for different frequencies. Thus we obtain a series of parallel lines (for different n) with the same slopes and independently of "n", the calculated $E_a$ value remains constant and equal to that determined from the maximum. This result shows that such a method of approach is applicable and we use it also to determine $E_a$ for the Y08-N sample. The definite value of $E_a$ at $H_{ac}$=7.95 a/m (0.1 Oe) is very low ($E_a$=0.04). Most probably it is related to intragranular flux pinning and corresponding intergranular value should be even smaller.

On Fig.3 the $1/T_p$ vs. ln(**f**) dependences are presented for all investigated samples. It is seen from the figure that underdoped samples (Y08-N and Gd123) have higher slope of the $1/T_p$ vs. ln(**f**) dependence and hence lower activation energy than the overdoped samples (Y123, Y08-48 and Y08-100). Increasing the level of overdoping raises the activation energy. The exact values of $E_a$ for $H_{ac}$=7.95 a/m (0.1 Oe) are given in Table 1.

On Fig.4 frequency dependences for Y123 and Y08-100 samples are presented at constant magnetic field amplitude $H_{ac}$=7.95 a/m (0.1 Oe). This comparison is important in order to establish the exact role of Ca substitution, responsible for higher level of overdoping. We have to draw attention to two facts. First, Ca substitution reduces the width of intergranular transition. Observation of intergranular transition in $\chi'(T)$ for granular superconductors is a result of coupling of grains via Josephson currents through insulating barriers (unfavorably oriented grains, oxygen–depleted GB) [22]. The observed narrowing of intergranular transition could be a result of improvement of intergrain connections and intergranular resistance reduction [23]. All this is in consent with Ca segregation in GB regions, reported by different authors [8, 24] Second, the intergranular transition shift, with increasing the frequency, is smaller in Ca substituted sample than in non substituted one. Higher level of overdoping leads to a higher intergranular current, supporting intergranular flux pinning in Ca substituted sample.

In order to establish whether there exists different vortex dimensionality in underdoped (Y08-N) and overdoped (Y08-100) samples we performed experiments for the determination of activation energy at four different AC field amplitudes-7.95 a/m (0.1 Oe); 79.57 a/m (1 Oe); 795.78 a/m (10 Oe) and 1591.59 a/m (20 Oe). On Fig.5 (a, b) the $E_a(H_{ac})$ dependences are presented for both samples, respectively. Magnetic field dependence of the activation energy displays a logarithmic behavior in 2D vortex system [25]. For underdoped Y08-N sample the $E_a(H_{ac})$ dependence is found to be logarithmic. This correlates with the small carrier concentration, large **c** lattice parameter (11.818 A) and high anisotropy of this sample. In spite of the fact that this result probably refers to the intragranular pinning it is reasonable to expect similar behavior for intergranular pinning also. For an overdoped Y08-100 sample the experimentally found points are well approximated by the $E_a \sim H_{ac}^{-2/3}$ dependence shown in the inset of Fig.5b. According to [26] this dependence is characteristic of a 3D vortex system.

C. Flux line dynamical regimes

AC magnetic susceptibility measurements are used for studying the different dynamical regimes of the flux lines. The response of HTS to an AC magnetic field can be linear or non-linear. The linear response can be divided into two different regimes: TAFF and flux flow (FF). In the TAFF regime vortex are thermally activated to jump between different metastable states and contribute to the AC response [27]. Viscous motion of the vortex liquid dominates the response in the FF regime. In both regimes the linear response is due to a resistive state E=ρJ in the superconductor and ρ(B,T) is independent of J. In the non-linear flux creep (FC) regime sample's resistivity is J dependent and a non-zero third harmonic signal appears.

In order to compare intergranular behavior in underdoped and overdoped samples, which have different $T_{cinter}$ and different geometry we used the Coles-Coles ($\chi'$ versus $\chi''$) plot of magnetic susceptibility. The advantage of this presentation is the elimination of the geometry dependent factors [28, 29] as well as the possibility to compare different samples without knowing the exact J(T) dependence [28]. On Fig.6 (a, b) Coles-Coles plots for underdoped Gd123 and overdoped Y08-100 samples are presented, respectively. The values of abscissas are normalized by using the maximum value of $|\chi'_1|$ for the corresponding sample, where the Meissner state is reached. $\chi''_1$ is also normalized to the corresponding maximum value. It has been previously shown [28] that this presentation gives the same results when the fundamental susceptibility is measured as a function of temperature at constant magnetic field or the measurement is performed at varying magnetic fields and constant temperature. We also established this experimentally. In Fig.6 experimental results of type $\chi_1(T)_{Hac=const}$ are used and initial AC susceptibility data for investigated samples are presented at insets. For underdoped Gd123 sample (Fig.6a) two dependences are presented for $H_{ac}$=7.95 a/m (0.1 Oe) and $H_{ac}$=39.78 a/m (0.5 Oe), f=1048 Hz.

It is seen from the Coles-Coles plots that the maximum is shifted to the (-1) side of $\chi'_1$ axis. According to Shantsev et al [30] the steeper slope at $\chi' \to$ -1 and maximum shift to the negative -1 side is the indication of flux creep existence in the sample. For overdoped sample Y08-100 the presented curves (Fig.6b) are measured for more stronger magnetic fields $H_{ac}$=79.5 a/m (1 Oe) and $H_{ac}$=1.59.10$^3$ a/m (20 Oe). In both fields the maximum occurs at $\chi'_1/\chi'_{1\ max}$ =0.34 which is very close to the value 0.38 predicted from Bean critical state model [30]. Therefore, $J_c$ is almost independent on magnetic field in overdoped sample up to the highest field used in the investigations, while for underdoped sample the flux creep is present even in lower magnetic fields.

D. Temperature dependence of critical current density

Maximum of $\chi''_1(T)$ is reached when the applied magnetic field penetrates to the center of the sample. The increase of the magnetic field amplitude shifts the maximum position towards lower temperatures [14]. This can be used for determination of J(T) dependence when the sample's shape is known. The samples have been approximated with long cylinders (L=10R, where R is the radius and L is the length of cylinder) and the critical current density at $T_p$ has been calculated according to a relation $J(T) = H_{ac}/R$. The $\chi''_1(T)_{Hac=const}$ dependences at four different $H_{ac}$ amplitudes indicated above are used for $J_c(T)$ determination in underdoped Gd123 and overdoped Y08-48, Y08-100 samples. The corresponding $J_c(T)$ dependences are presented on the Fig.7(a, b). The results are well approximated (with small square deviation) with linear fit for underdoped and quadratic fit for overdoped samples. Similar behavior was already observed from direct $J_c(T)$ measurements in series of Ca substituted $GdBa_2Cu_3O_z$ samples [31]. The $J_c(T)$ dependence for underdoped sample is understandable. Within the individual grain $CuO_2$ planes are separated from insulating layers. In the frame of polycrystalline sample large angle grain boundaries depleted of carriers or infavourably oriented grains behave as Josephson junctions. Thus the observed linear temperature dependence of $J_c$ may be ascribed to the S-I-S type joints existence in grains and between them. More unusual is quadratic $J_c(T)$ dependence for overdoped sample, which is an indication of S-N-S type joints presence. Similar behavior is well known for superconducting samples with various metallic additions (Ag, Pt) [32]. Using this analogy we can suppose that normal carriers and/or small regions of normal state are present at grain boundaries of overdoped samples. It is known that Ca segregates at the grain boundary regions in substituted Y123 phase [8, 24, 33]. Ca substitution is a structural imperfection leading to the structure distortion not only in grains but at the grain boundaries also due to segregation. We can speculate that deformed regions with suitable dimensions could act as pinning centers increasing the intergranular pinning energy in overdoped samples. On the other side Y08-48 and Y08-100 specimens consists of Ca substituted regions with dimensions comparable to the unit cell with different oxygen content and different $T_c$. At the conditions of experiment Ca substituted regions become normal at temperatures lower than the rest and act as pinning centers. This can explain the observed S-N-S type behavior of $J_c(T)$ also.

In fact we established that a correlation exists between the intergranular critical current and flux pinning activation energy. In underdoped samples the intergranular current shows S-I-S behaviour and the activation energy is small, while in overdoped samples the intergranular current is changed to S-N-S type and activation energy increases. Recently Chen et al [34] reported also that the maximum Josephson current across the grains determines the intergranular flux behaviour. It was found out that Ca substitution not only increases carrier concentration and improves intergranular current but it also has the significance for intra- [35] and intergranular pinning.

In conclusion, we have demonstrated that a close relation exists between the intergranular critical current and intergranular vortex dynamics. Overdoped samples have higher activation energy for TAFF and its intergranular critical current is governed by S-N-S type connections. Underdoped samples show low activation energy and S-I-S type intergranular current. 2D pancake vortices are characteristic of underdoped samples, while 3D vortices exist in overdoped samples. As a result in underdoped samples already small magnetic fields excites flux creep regime in the intergranular region. Even twenty times higher magnetic field is not high enough to transfer the overdoped sample into flux creep regime.

In fact we demonstrate that Ca substitution not only increases carrier concentration, but improves intergranular flux pinning too.

## Acknowledgements

This work is carried out in the framework of Polish- Bulgarian Interacademic cooperation scheme and FU07-CT-2007-00059. The authors would like to thank Prof. V. Kovachev for careful reading the manuscript and useful remarks.

**Figure Captions**

Fig.1 X-ray diffractograms for all investigated samples.

Fig.2 Temperature dependences of (a) real and (b) imaginary parts of fundamental AC magnetic susceptibility for sample Y08-N at $H_{ac}$=7.95 A/m (0.1 Oe) and indicated frequencies.

Fig.3. $1/T_p$ vs. ln(f) dependences for all investigated samples. The magnetic field amplitude is 7.95 A/m (0.1 Oe) and frequency is changed in the interval 145 – 5.3.$10^3$ Hz.

Fig.4. Temperature dependences of fundamental AC magnetic susceptibility for Ca substituted (Y08-100) and non substituted (Y123) samples at constant magnetic field amplitude 7.95 A/m (0.1 Oe) and different frequencies.

Fig.5a. Activation energy ($E_a$) vs. magnetic field amplitudes for underdoped sample Y08-N.

Fig.5b. Activation energy ($E_a$) vs. magnetic field amplitudes for overdoped sample Y08-100. Inset is presented $E_a$ as a function of $H_{ac}^{-2/3}$.

Fig.6a. Coles-Coles plots for uderdoped Gd123 sample. The initial AC susceptibility data are presented inset

Fig.6b. Coles-Coles plots for overdoped Y08-100 sample. The initial AC susceptibility data are presented inset.

Fig.7. Temperature dependences of critical current density for (a) underdoped Gd123 and (b) overdoped (Y08-48; Y08-100) samples.

Table 1. Samples composition, symbols, lattice parameters, intra- and intergranular critical temperatures and activation energy for TAFF.

| Sample | Symbol | a (A) | b (A) | c (A) | $T_{c,intra}$ (K) | $T_{c,inter}$ (K) | $E_a$ (eV) |
|---|---|---|---|---|---|---|---|
| $Y_{0.8}Ca_{0.2}Ba_2Cu_3O_z$ | Y08-N | 3.849 | 3.849 | 11.818 | ~50.0 | - | 0.004 |
| $Gd_1Ba_2Cu_3O_z$ | Gd123 | 3.843 | 3.906 | 11.718 | 91.7 | 85.8 | 0.397 |
| $Y_1Ba_2Cu_3O_z$ | Y123 | 3.828 | 3.846 | 11.642 | 92.0 | 90.1 | 1.118 |
| $Y_{0.8}Ca_{0.2}Ba_2Cu_3O_z$ | Y08-48 | 3.818 | 3.877 | 11.676 | 82.7 | 69.4 | 2.646 |
| $Y_{0.8}Ca_{0.2}Ba_2Cu_3O_z$ | Y08-100 | 3.814 | 3.868 | 11.666 | 81.5 | 66.6 | 2.837 |

Figure 1.

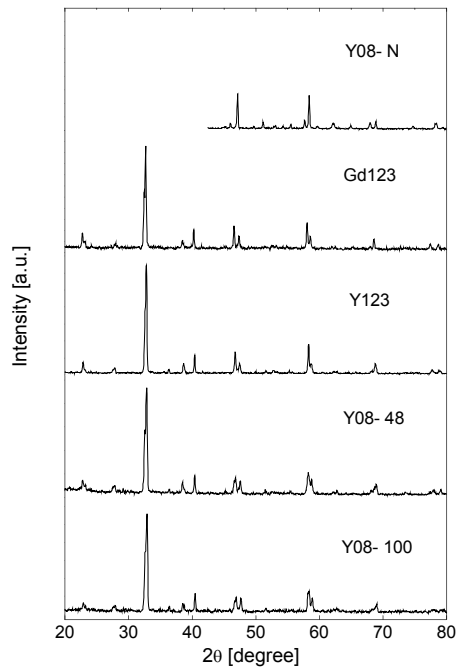

Figure 2 (a).

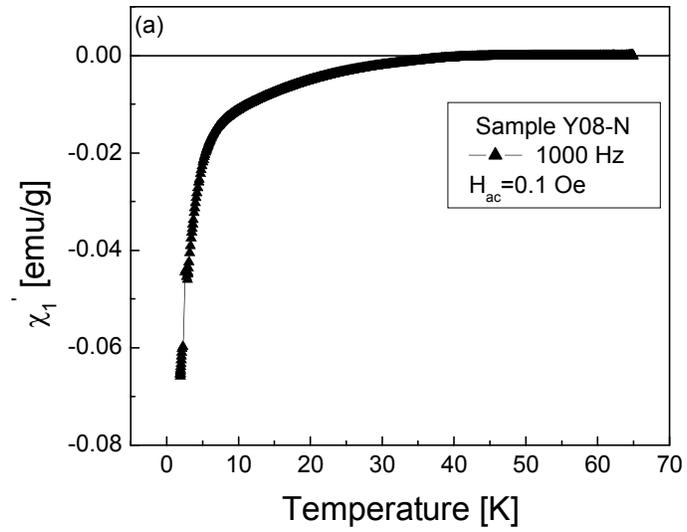

Fugure 2 (b).

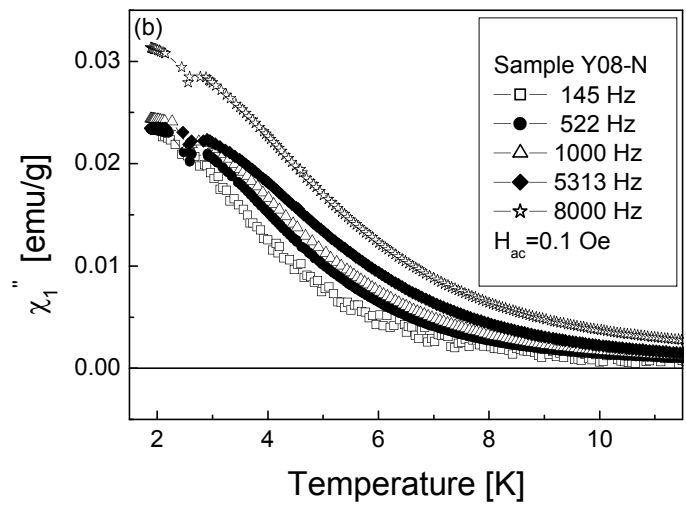

Fugure 3.

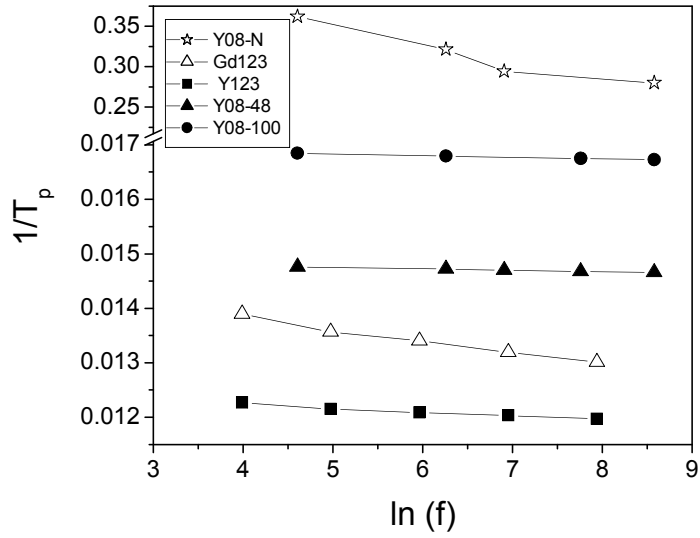

Figure 4.

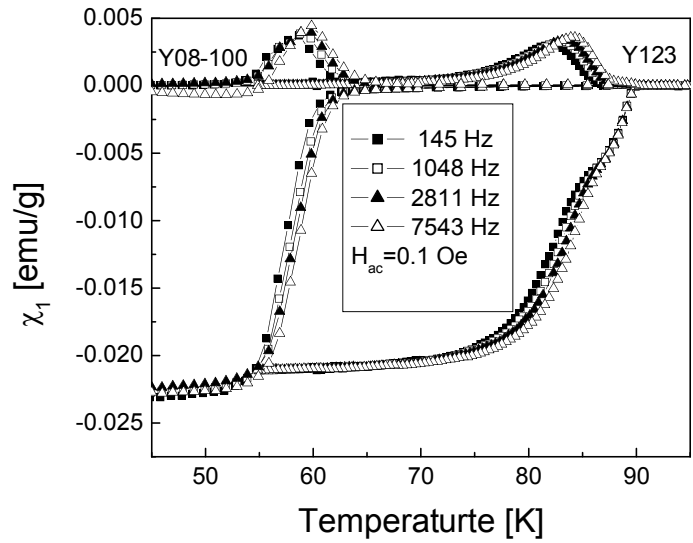

Figure 5 (a).

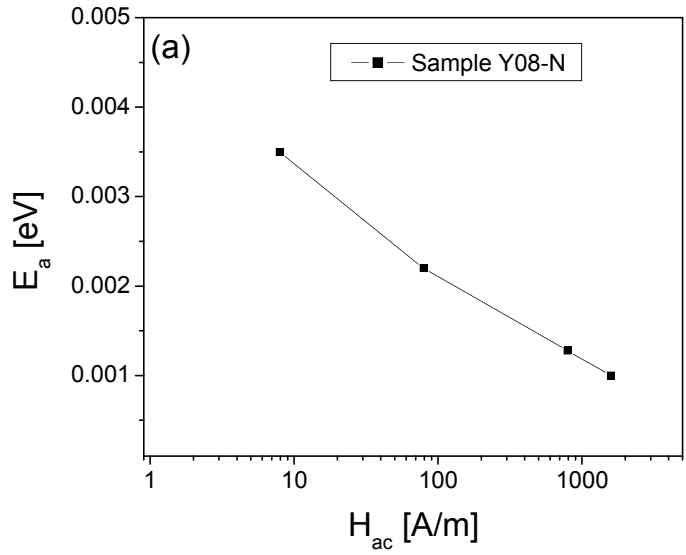

Figure 5 (b).

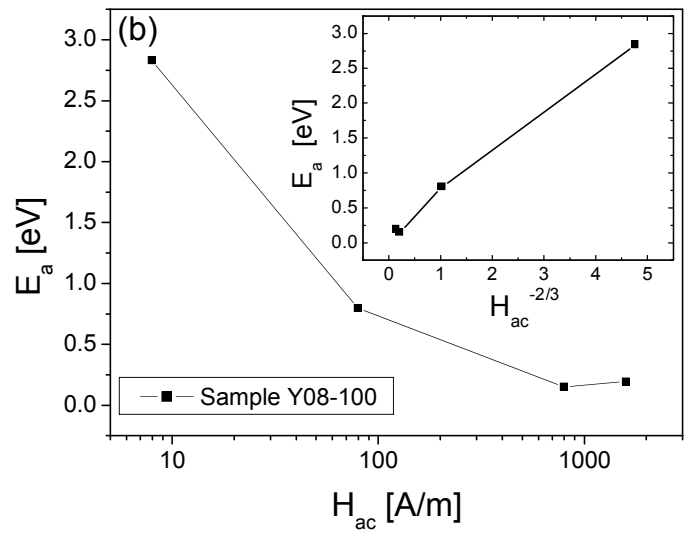

Figure 6 (a).

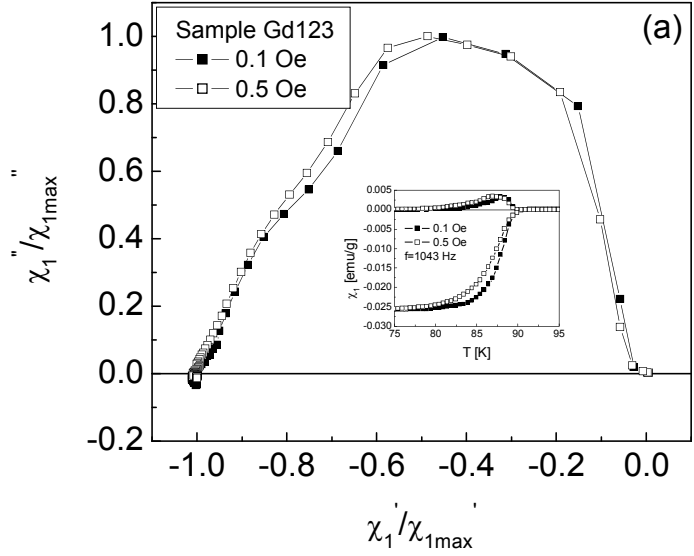

Figure 6 (b).

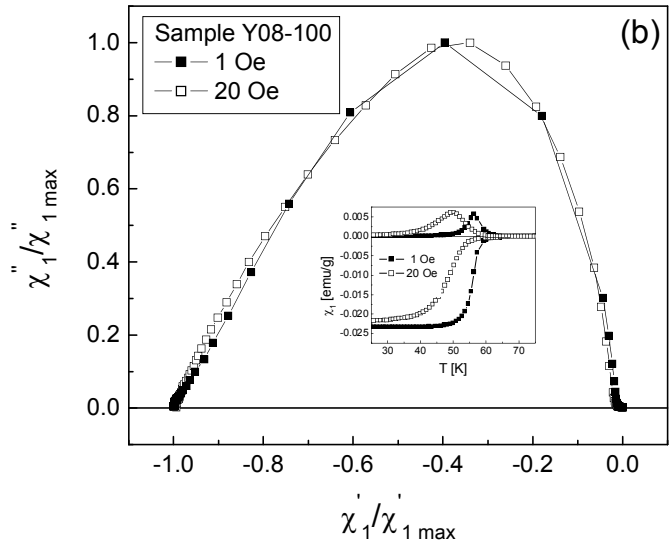

Figure 7 (a).

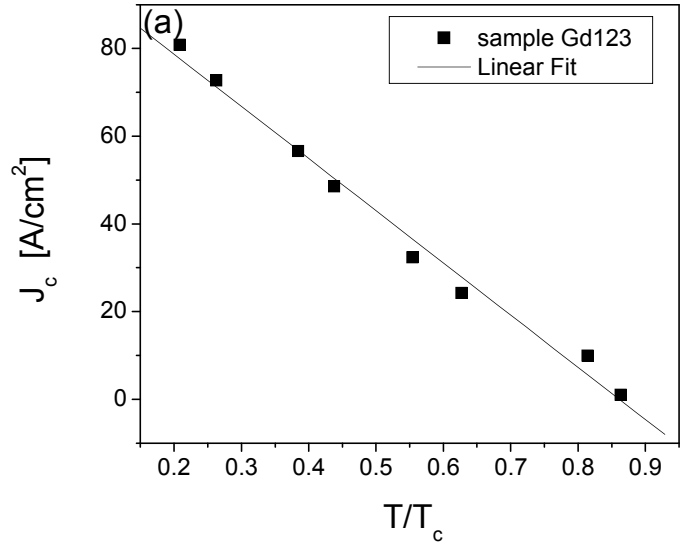

Figure 7 (b).

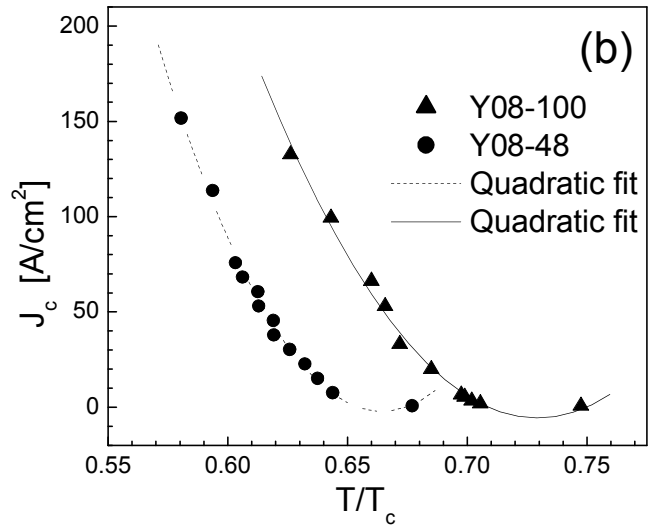